%% file: paper8-arxiv.tex
\documentclass[superscriptaddress,twocolumn,amsmath,amssymb,prl]{revtex4}
\pdfoutput=1

\usepackage{times}
\usepackage{graphicx}    
\usepackage{color}          

\bibpunct{}{}{,~}{s}{,}{,}

\include{defs_thesis}

\begin{document}

\title{Ultrafast entangling gates between nuclear spins using photo-excited triplet states}

\author{Vasileia Filidou\footnote{These authors contributed equally to this work}}
\author{Stephanie Simmons$^*$}
\affiliation{Department of Materials, University of Oxford, Parks Road, Oxford OX1 3PH, UK}
\author{Steven D. Karlen}
\affiliation{Department of Materials, University of Oxford, Parks Road, Oxford OX1 3PH, UK}
\affiliation{Department of Chemistry, University of Oxford, Oxford OX1 3TA, UK}
\author{Feliciano Giustino}
\affiliation{Department of Materials, University of Oxford, Parks Road, Oxford OX1 3PH, UK}
\author{Harry L. Anderson}
\affiliation{Department of Chemistry, University of Oxford, Oxford OX1 3TA, UK}
\author{John J. L. Morton}
\email{john.morton@materials.ox.ac.uk}
\affiliation{Department of Materials, University of Oxford, Parks Road, Oxford OX1 3PH, UK}
\affiliation{CAESR, The Clarendon Laboratory, Department of Physics, University of Oxford, OX1 3PU, UK}

\date{\today}

\begin{abstract}
The representation of information within the spins of electrons and nuclei has been powerful in the ongoing development of quantum computers~\cite{Cory19971634,Laflamme19981941}. Although nuclear spins are advantageous as quantum bits (qubits) due to their long coherence lifetimes (exceeding seconds~\cite{Ladd2005014401,morton:qmemory}), they exhibit very slow spin interactions and have weak polarisation. A coupled electron spin can be used to polarise the nuclear spin~\cite{DNP1,DNP2,simmons11} and create fast single-qubit gates~\cite{morton:bangbang, coryPRA2008}, however, the permanent presence of electron spins is a source of nuclear decoherence. Here we show how a \emph{transient} electron spin, arising from the optically excited triplet state of \csixty, can be used to hyperpolarise, manipulate and measure two nearby nuclear spins. 
Implementing a scheme which uses the spinor nature of the electron~\cite{Schaffry2011}, we performed an entangling gate in hundreds of nanoseconds: five orders of magnitude faster than the liquid-state J coupling.
This approach can be widely applied to systems comprising an electron spin coupled to multiple nuclear spins, such as NV centres~\cite{Robledo2011}, while the successful use of a transient electron spin motivates the design of new molecules able to exploit photo-excited triplet states.
\end{abstract}

\maketitle

Different quantum systems possess different advantages as qubits, stimulating the use of so-called hybrid approaches to quantum computing~\cite{mortonlovett}. Examples include interfacing superconducting qubits with spin ensembles~\cite{kubo2011}, optical photons with defects in solids~\cite{photonnv}, and electron spins with nuclear spins~\cite{morton:qmemory,nvmemory}. 
Spins controlled with nuclear magnetic resonance (NMR) have played an important role in the developement of much experimental work in quantum information processing, showcasing high-fidelity control~\cite{Laflamme2005}, complex demonstrations of quantum algorithms~\cite{Vandersypen2001} and many-qubit decoupling strategies~\cite{JonesOverview11}. 
Nuclei in such systems are only weakly coupled: the indirect J-coupling interaction available in liquid-state NMR can be on the order of 100~Hz~\cite{Vandersypen2001}, though nuclear spin dipole couplings in the solid state can exceed 10~kHz.~

This weak coupling places a lower limit on the duration of a quantum logic operation between two spins, and thus the computational speed of a nuclear spin-based quantum information processor.
Additionally, the weak magnetic moment of nuclear spins leads to a weak polarisation in general (typically less than 0.01$\%$ for liquid state NMR), making the scaling-up of the initial demonstrations very challenging unless a method for nuclear spin cooling can be applied~\cite{algorithmiccooling}. 

These limitations can be addressed by making use of a coupled electron spin. Highly polarised electron spin states can be transferred to the nuclear spin coherently using SWAP operations~\cite{daviesendor, morton:qmemory,simmons11}, or incoherently using a family of dynamic nuclear polarisation (DNP) methods~\cite{DNP1,DNP2}. Typical single-qubit gate times for electron spins are tens of nanoseconds, and given typical electron-nuclear couplings (in the range 1--100 MHz), it is possible to manipulate a nuclear spin on these timescales. For purely isotropic couplings, phase gates can be applied to nuclear spin qubits to perform dynamic decoupling~\cite{morton:bangbang,tyryshkin:siqubits}, while using anisotropic coupling, more general gates have been applied  to single nuclear spins~\cite{khaneja2007PRA,mitrikasPRA2010, coryPRA2008}, or, recently, two nuclear spins~\cite{zhang11}.

A disadvantage of using coupled electron spin is that the nuclear spin coherence time can be strongly limited by electron spin relaxation or flip-flop processes~\cite{morton:qmemory}. A better strategy invokes an electron spin only at certain key times, for example to hyperpolarise the nuclear spins or to perform fast logic gates, so that there is minimal long-term impact on nuclear decoherence\cite{gauger2008}.  

\begin  {figure}[t]
\begin  {center}
\includegraphics[width=8.5cm]{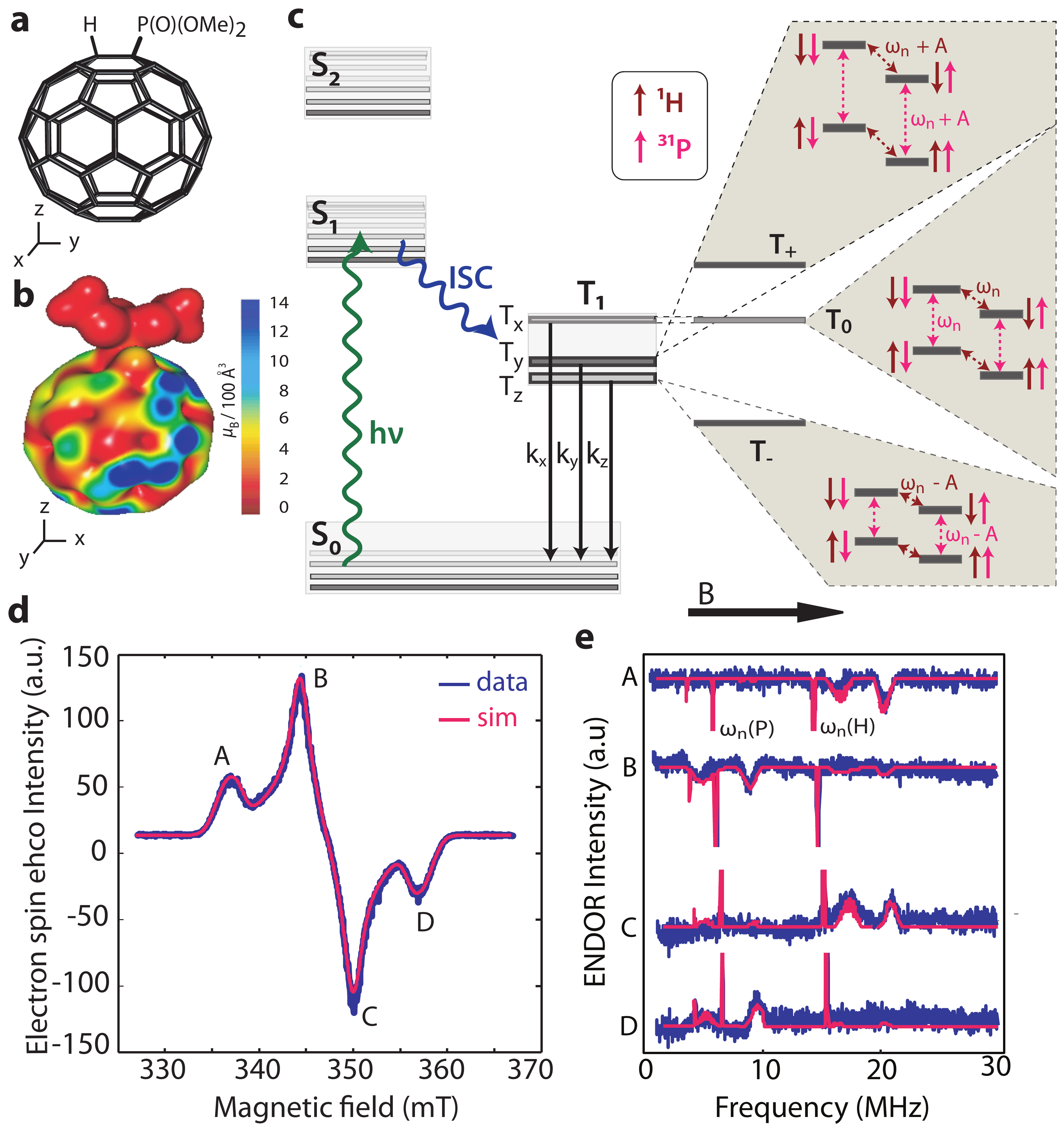}
\caption{\textbf{%Dimethyl fullerene phosphonate (DMHFP) 
The DMHFP molecule has an excited triplet state which couples to two nuclear spins: \hone~and \ptone.}
a) Illustration of the DMHFP molecule. 
b) Spin density distribution in the excited triplet state T$_1$, where blue areas correspond to atoms with high electron spin density.
c) The system has a diamagnetic singlet ground state S$_0$ which can be excited through ISC to a triplet state, T$_1$. The application of a magnetic field mixes the triplet sublevels in an orientation dependent manner. The triplet further couples to the two nuclear spins directly bonded to the cage, causing further splittings due to nuclear Zeeman and hyperfine energies.
d) An electron spin echo-detected field sweep showing the ESR spectrum of the triplet state. Resonances at different fields correspond to different molecular orientations with respect to the applied field.
e) Electron nuclear double resonance spectroscopy (ENDOR) is applied in four field positions in order to extract the isotropic hyperfine interaction between the triplet electron spin and the two nuclear spins \hone~and \ptone, measured to be 6~MHz and 11~MHz respectively).
}
\label{fig:one}
\end {center}
\end {figure}

To explore such possibilities, we synthesised the fullerene derivative DMHFP~\cite{dmfphsynth}, illustrated in \Fig{fig:one}a, containing two nuclear spins (\ptone~and \hone) which are directly bonded to a \csixty~fullerene cage. The molecule has a  diamagnetic singlet ground state which can be photo-excited to populate the first excited single state. This state undergoes intersystem crossing (ISC) to a long-lived triplet state which is paramagnetic ($S=1$) with electron spin density delocalised over the cage (see \Fig{fig:one}b,c).  This system provides all the ingredients to explore nuclear spin manipulations mediated by a transient electron spin using a combination of optical excitation, electron spin resonance (ESR) and NMR control.

We begin by characterising the spin Hamiltonian of the DMHFP molecule:

\begin{align}
\mathcal{H}=&\mu_B\vec{S}{\bf g_e}\vec{B}+\vec{S}{\bf D} \vec{S}+ \sum_{i= \trm{\ptone, \hone}}\vec{S}{\bf A}(i) \vec{I}_i \\ \nonumber
&  + J I_{1,z} I_{2,z} + \gamma_{i,n}\vec{I}_i \cdot\vec{B}
\end{align}
where 
$\vec{B}$ the applied magnetic field, $\gamma_{i,n}$ the nuclear gyromagnetic ratio, ${\bf g_e}$ the electron  ${\bf g_e}$-factor tensor, $\mu_B$ the Bohr magneton, {\bf D} the zero-field splitting (ZFS) tensor for the $S=1$ triplet state, {\bf A}(i) the hyperfine coupling tensor between the triplet and the nuclear spins $i$, and $J$ is the coupling between the nuclear spins. All terms involving $\vec{S}$ vanish in the electronic ground state. 

By performing pulsed ESR immediately following a 532~nm laser pulse we examine the properties of the triplet state. \Fig{fig:one}d shows the intensity of an electron spin echo as a function of the applied magnetic field at X-band (9.7~GHz microwave frequency). 
By comparing the spectrum to simulations~\cite{easyspin}, we obtain the ${\bf g_e}$-tensor and the principal values of the ZFS tensor $D_{xx}=52.13,  D_{yy}= 159.53, D_{zz} = -211.66~$MHz. These ZFS parameters depend on the spatial distribution of the triplet wavefunction and characterise the strength and the asymmetry of the electron dipolar coupling.

The hyperfine coupling between the triplet electron spin and the $^{1}$H and $^{31}$P nuclear spins can be measured using the Davies electron-nuclear double resonance (ENDOR) method~\cite{daviesendor}, applied to select different molecular orientations within the sample. The spectra, shown in \Fig{fig:one}e, show narrow peaks around 6 and 14~MHz corresponding to the nuclear Larmor frequencies of the  $^{31}$P and $^{1}$H spins, and arising from nuclear transitions in the T$_0$ subspace where the hyperfine coupling is negligible (see \Fig{fig:one}c). The ENDOR efficiency of these peaks is 90--100$\%$ enabling high-fidelity control on these transitions.

The other peaks in the ENDOR spectra arise from the (orientation-dependent) hyperfine coupling in the T$_{\pm}$ subspaces. Fitting yields the isotropic hyperfine coupling terms A(\hone) = 6.0~MHz and A(\ptone) = 11~MHz, consistent with DFT modelling (see Supporting Information). This hyperfine coupling allows for conditional electron/nuclear spin operations to be performed, however the breadth of these peaks (arising from the randomly oriented solid) results in poor fidelity nuclear spin control in the T$_{\pm}$ subspaces. 
Nevertheless, we will show that is is possible to apply entangling operations in the T$_0$ subspace where there is negligible coupling between the nuclear and electron spins.

\begin  {figure}
\begin  {center}
\includegraphics[width=8.5cm]{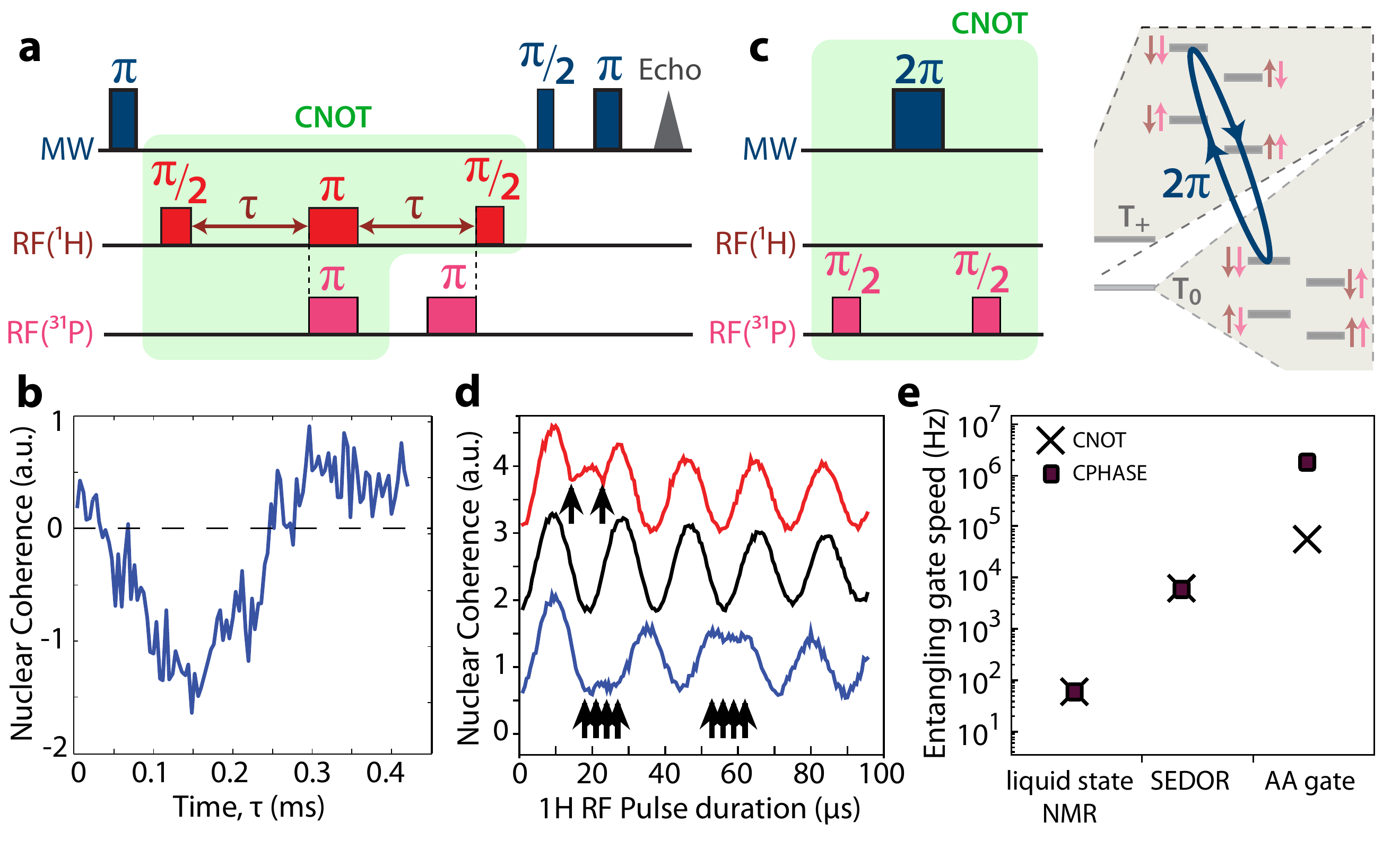} 
\caption{
\textbf{Two different ways to implement nuclear spin entangling gates using the triplet electron spin.}
a) A technique based on the spin echo double resonance (SEDOR) sequence can be used to measure the dipolar coupling between spins. The sequence resembles a Hahn-echo experiment on one spin (\hone), however both spins are flipped during the refocussing pulse such that the sign of their coupling remains unchanged. Microwave pulses before and after are used to prepare and measure the \hone~nuclear spin coherence. 
b) The SEDOR sequence produces an oscillation in the nuclear coherence as $\tau$ is swept, corresponding to a nuclear coupling of 3kHz. 
c) An alternative implementation of a CNOT gate consists of two Hadamard gates ($\pi$/2 pulses) and a CPHASE gate created by applying a selective 2$\pi$ pulse to the electron spin.
d) An ultrafast nuclear phase gate (created by a 2$\pi$ microwave pulse) is applied during nuclear spin Rabi oscillations on the \hone~spin at points marked with arrows.
e) Comparison of the speed of the entangling operations. From the liquid state to the photo-excited solid state with the application of CPHASE gates there is an improvement of the entangling speed of five orders of magnitude. }
\label{fig:gates}
\end {center}
\end {figure}

NMR experiments in the absence of optical excitation reveal the ground-state $J$-coupling between \ptone~and \oneh~to be 30~Hz, which leads to a nuclear controlled-NOT (CNOT) operation time of 17~ms. In the solid state the dipolar coupling can be measured using the spin echo double resonance (SEDOR) pulse sequence~\cite{sedor, schweiger01}, shown in~\fig{fig:gates}a. SEDOR can be interpreted as a standard NMR CNOT operation modified to allow for initialisation and readout by the electron spin. The timing of the refocussing pulses is swept to identify the CNOT coupling time of 160~\mus, corresponding to a 3~kHz nuclear coupling.
By comparing this coupling time to the lifetime of the triplet state (approximately 0.5~ms) and nuclear \ttwo~times (0.20(4) and 1.9(4)~ms for \hone~and \ptone~respectively), we see that shorter entangling gate times are needed for higher fidelity operations.
This can be achieved by using a triplet electron spin transition to apply an Aharanov-Anandan (AA)~\cite{AharonovAnandan,morton:bangbang} controlled-phase (CPHASE) gate to the nuclear spins, as proposed in Ref 10.%~

If a quantum state is taken through a closed-loop trajectory in Hilbert space, it acquires a geometric phase equal to half the solid angle mapped out by that trajectory. 
Because the field is on-resonance with an electronic transition corresponding to a particular configuration of the \hone~and \ptone~nuclear spins, we can apply a Toffoli gate: a microwave pulse which only rotates the electron spin when the nuclear spins are, say, in the $\ket{\uparrow\uparrow}$ state. Applying a 2$\pi$ pulse to the electron in this way imparts a $\pi$ phase shift to the $\ket{\uparrow\uparrow}$ state, with respect to the others in the T$_0$ subspace, equivalent to a CPHASE operation acting upon an effective nuclear spin basis. Both the CPHASE and CNOT operations are well-defined with respect to an effective eigenbasis where $\ket{4}$ is the nuclear spin eigenstate associated with the chosen electronic transition. The eigenstates $\ket{2}$ and $\ket{3}$ differ from $\ket{4}$ by a \hone\ and \ptone\ spin flip, respectively, and $\ket{1}$ differs from $\ket{4}$ by a flip of both nuclear spins. 

The duration of the CPHASE operation is limited only by the hyperfine coupling strength, such that a CPHASE gate on the timescale of hundreds of nanoseconds can be performed. To illustrate how this phase gate can be applied to the individual nuclear spins, we applied $2\pi$ microwave pulses while driving nuclear Rabi oscillations (see \fig{fig:gates}d). We verified that this CPHASE behaviour was conditional by observing uninterrupted Rabi oscillations on the complementary nuclear subspace. A CNOT operation can be built by combining the CPHASE gate with Hadamard rotations (see \Fig{fig:gates}c).

\begin  {figure}
\begin  {center}
\includegraphics[width=8.5cm]{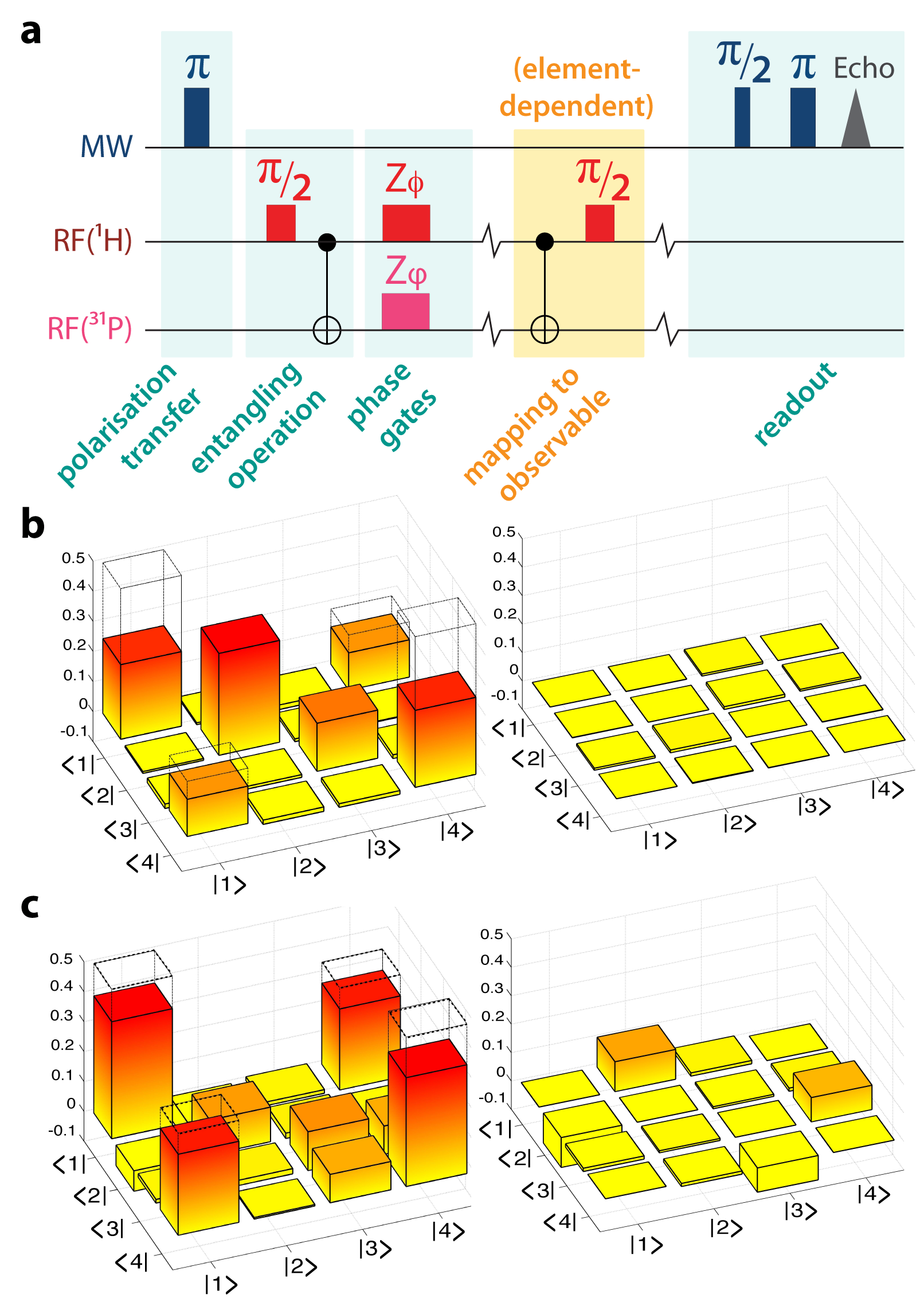} 
\caption{{\bf Density matrix tomography results.} 
a) General pulse sequence for the extraction of the $\ketbra{1}{4}, \ketbra{4}{1}$ elements of the density matrix. The CNOT representations correspond to the entangling gates as represented by the shaded areas of \Fig{fig:gates}a and \Fig{fig:gates}b. The pulses mapping a density matrix element to the observable changes depending upon the density matrix element. b) Density matrix obtained using the coupling-based entangling gates. The slow coupling leads to a fidelity of the operations of 34$\%$. c) Density matrix obtained using CPHASE-based entangling gates. The ultrafast CPHASE gate entangles the nuclear spins in 220~ns and the fidelity increases to 65$\%$. }
\label{fig:denmat}
\end {center}
\end {figure}

In order to compare the performance of these two entangling CNOT operations, we attempt to put the two nuclear spins into a Bell state and then perform tomography of the effective spin density matrix, building upon methods described elsewhere~\cite{simmons11} (see Supplementary Information). In short, each element of the density matrix must be mapped in turn onto the observable electron spin transition, through a combination of microwave and RF pulses. Notably, a CNOT (or similar) operation is needed when reading the zero- or double-quantum coherences, and these can be accomplished using either of the methods introduced above. To improve the fidelity of the tomography, each element of the density matrix is imprinted with a particular time-varying phase applied to the nuclear spins using RF pulses. The total pulse sequence for generating the pseudo-entangled state and measuring it using quantum state tomography is given in \fig{fig:denmat}a.

In \Fig{fig:denmat}b and \Fig{fig:denmat}c we compare the density matrices obtained using the two different implementations of the CNOT gate: 1) exploiting the nuclear dipolar coupling in the solid state, or 2) the triplet-mediated AA CPHASE operation combined with Hadamard gates. The fidelities of the final density matrices $\rho_\textrm{D}$ with respect to the ideal Bell state $\rho_\textrm{B}$, calculated according to $F(\rho_\textrm{B},\rho_\textrm{D}) = \left( \textrm{Tr} \left( \sqrt{ \sqrt{\rho_B} \rho_D \sqrt{\rho_B} }\right) \right)^2$ are 34$\%$ and 65$\%$ respectively.
The increased fidelity of the latter approach is due primarily to the much shorter gate times: the CPHASE entangling gate is performed in only 220~ns, and adding the Hadamard gates yields a CNOT gate time of 34.2~$\mu$s. In comparison, the CNOT based on the dipolar coupling had a duration of 160~$\mu$s. 
The maximum fidelity of each approach given the finite triplet recombination time for this molecule is 68$\%$ and 85$\%$, respectively. The residual imperfection is due to the limited fidelity of the CPHASE operation, as evidenced in \Fig{fig:gates}d, and small gate imperfections, consistent with simulations incorporating a 4\% error on each nuclear gate in the sequence.

The lifetime of the double quantum coherence $T_{2,\trm{DQC}} \approx 100~\mu$s and the lifetime of the zero quantum coherence $T_{2,\trm{ZQC}} \approx 200~\mu$s which indicates that they are limited by the hydrogen $T_2$ in the photo-excited state. Both nuclear coherence lifetimes in the liquid state are longer by orders of magnitude, which motivates the further study of controlled de-excitation of the triplet state in place of the stochastic decay process to remove the electron as a source of nuclear decoherence. 

Although in disordered frozen solutions some limited orientation selection is possible, the use of a single crystal sample would improve the uniformity and hence fidelity of the entangling operations. Due to the orientation-dependent triplet populations, a single-crystal sample would additionally allow for the selection of an orientation with the highest electron triplet polarisation, maximising the quality of entanglement. This technique is readily extensible to a wide range of molecular systems where a large number of nuclear spins are coupled to photoexcited triplet states, which can in turn be placed into arrays.

\section{Methods Summary}

Dimethyl [9-hydro(C$_{60}$-I$_h$)[5,6] fulleren-1(9H)-yl] phosphonate was prepared following the procedure reported by Nakamura et al., Scheme 1 \cite{dmfphsynth}. Mono-functionalization of C$_{60}$ was performed using dimethyl phosphonate in a solution of toluene and HMPA at 120$^\circ$C in the presence of oxygen. The product was purified by silica column chromatography (toluene and
ramped to 10\% ethyl acetate in toluene. 

Pulsed electron spin paramagnetic experiments were performed using an X-band (9--10 GHz) Bruker Elexsys680 spectrometer equipped with a low-temperature helium-flow cryostat (Oxford CF935). The arbitrary phase radiofrequency pulses were generated using a using a Rohde and Schwarz AFQ100B together with an Amplifier Research 500W amplifier. 
Photo-excitation was achieved using an Nd-YAG laser at 532 nm with a 10 Hz repetition rate, using 10~mJ pulses, 7ns in length.
Microwave pulse lengths were 128 ns for $\pi/2$ pulses, and 220 ns for both $\pi$ and $2\pi$ pulses. The duration of RF pulses (both $\pi/2$ and $\pi$) was 17 $\mu$s. The samples were prepared in toluene-d8 with concentration of  $4\times10^{-4}$ M, deoxygenated and flame sealed under vacuum and flash frozen in liquid nitrogen.

\begin{acknowledgments}
\textit{Acknowledgements} : We thank 
Brendon Lovett, Marcus Schaffry, Erik Gauger, Chris Kay, Arzhang Ardavan, Andrew Briggs and Davide Ceresoli for helpful discussions. This work was supported by the EPSRC through CAESR (EP/ D048559/1) and the Materials World Network (EP/I035536/1), as well as by the European Research Council under the European Community's Seventh Framework Programme (FP7/2007-2013) / ERC grant agreement no.  279781. We thank the Violette and Samuel Glasstone Fund, Clarendon Fund, John Templeton Foundation, St John's College, Oxford, and the Royal Society for support.
\end{acknowledgments}

\bibliography{bib}

\end{document}

%% file: defs_thesis.tex
\newcommand {\trm}[1] {\textrm{#1}}

\newcommand {\Fig}[1] {Figure~\ref{#1}}
\newcommand {\fig}[1] {Figure~\ref{#1}}   % references Figure x
\newcommand {\ketbra}[2] {|#1 \rangle\langle #2|}

\newcommand{\be}{\begin{eqnarray}}
\newcommand{\ee}{\end{eqnarray}}
\newcommand{\bmat}{\begin{pmatrix}}
\newcommand{\emat}{\end{pmatrix}}

\newcommand{\csixty}{C$_{60}$}

\newcommand{\mus}{$\rm{\mu}$s}

\newcommand{\ptone}{$^{31}$P}
\newcommand{\hone}{$^{1}$H}

\newcommand{\oneh}{$^{1}$H}

\newcommand{\beq}{\begin{equation}}
\newcommand{\eeq}{\end{equation}}
\newcommand{\beqa}{\begin{eqnarray}}
\newcommand{\eeqa}{\end{eqnarray}}

\newcommand{\ket}[1]{\left| #1 \right\rangle}

\newcommand{\ttwo}{$T_2$}